\documentclass[onecolumn,showpacs,amsmath,amssymb,prd,nofootinbib]{revtex4-2}
\usepackage{graphicx}
\usepackage{epsfig}
\usepackage{enumerate}

\def\beq{\begin{equation}}
\def\eeqno#1{\label{#1}\end{equation}}

\def\rarrow{\rightarrow }

\def\dleft{\rlap{{\it D}}\raise 8pt
\hbox{$\scriptscriptstyle\Leftarrow$}}
\def\dright{\rlap{{\it
D}}\raise 8pt\hbox{$\scriptscriptstyle\Rightarrow$}}

\def\msun{M_{\odot}}
\def\az{a_{0}}
\def\azs{a_{0}^2}

\def\l0{\ell_{0}}

\def\rar{\rightarrow}
\def\s{\sigma}

\def\a{\alpha}

\def\l{\lambda}

\def\f{\phi}

\def\r{\rho}

\def\n{\nu}
\def\z{\zeta}

\def\A{\mathcal{A}}

\def\L{\mathcal{L}}

\def\SS{\mathcal{S}}

\def\Om{\Omega}

\def\d{\delta}

\def\drt{d^3\vr}
\def\a{\alpha}
\def\xlimin{{x\rarrow\infty \atop{\raise 1pt\hbox to 30pt
{\rightarrowfill}}}}
\def\limlim#1#2{{#1\rarrow #2 \atop{\raise 1pt\hbox to 30pt
{\rightarrowfill}}}}

\def\vr{{\bf r}}

\def\vv{{\bf v}}

\def\vg{{\bf g}}

\def\vF{{\bf F}}

\def\S{\Sigma}

\def\grad{\vec\nabla}
\def\div{\vec \nabla\cdot}
\def\gf{\grad\phi}

\def\Q{\mathcal{Q}}
\def\P{\mathcal{P}}
\def\L{\mathcal{L}}

\def\gh{g^{1/2}}

\def\a{\alpha}

\def\n{\nu}

\def\rh{\hat\rho}

\def\_#1{_{\scriptscriptstyle #1}}
\def\^#1{^{\scriptscriptstyle #1}}
\def\baz{\bar a_0}

\def\rp{\r\_p}

\def\oot{\frac{1}{2}}

\def\tpg{2\pi G}
\def\fpg{4\pi G}
\def\epg{8\pi G}

\def\gps{\grad\psi}
\def\hr{\hat\r}

\def\TP{\textbf{\textsf{P}}}
\def\Pss{\textsf{P}}
\begin{document}
\title{Generalizations of quasilinear MOND (QUMOND)}

\author{Mordehai Milgrom}
\affiliation{Department of Particle Physics and Astrophysics, Weizmann Institute}

\begin{abstract}
I present a class of theories that generalize quasilinear MOND (QUMOND).
These generalized-QUMOND (GQUMOND) theories are derived from an action, and, like QUMOND, they require solving only the linear Poisson equation (twice), and are thus amenable to relatively simple numerical solution. Unlike QUMOND, their Lagrangian depends on higher derivatives of the Newtonian potential. They thus dictate different ``phantom'' densities as virtual sources in the Poisson equation for the MOND potential.
These theories are not necessarily more appealing than QUMOND itself from a fundamental viewpoint. But, they might open new avenues to more fundamental theories, and, in the least, they have much heuristic value.
Indeed, I use them to demonstrate that even within limited classes of modified-gravity formulations of MOND, theories can differ substantially on lower-tier MOND predictions, showing that it would be imprudent, at present, to equate MOND itself with some specific effective theory, such as QUMOND.
Such GQUMOND theories force, generically, the introduction of dimensioned constants other than the MOND acceleration, $\az$, such as a length, a frequency, etc. As a result, some of these theories reduce to QUMOND itself only, e.g., on length scales (or, in other versions, dynamical times) larger than  some critical value. But in smaller systems (or, alternatively, in ones with shorter dynamical times), MOND effects are screened, even if their internal accelerations are smaller than $\az$.
In such theories it is possible that MOND (expressed as QUMOND) applies on galactic scales, but its departures from Newtonian dynamics are substantially suppressed in some subgalactic systems -- such as binary stars, and open, or globular star clusters. The same holds for the effect of the galactic field on dynamics in the inner Solar System, which can be greatly suppressed compared with what QUMOND predicts. Tidal effects of a galaxy on smaller subsystems are the same as in QUMOND, for the examples I consider. I also describe briefly versions that do not involve dimensioned constants other than $\az$, and yet differ from QUMOND in important ways, because they revolve around more than one acceleration variable.

\end{abstract}
\maketitle

\section{Introduction}
When applying modified Newtonian dynamics (MOND) \cite{milgrom83} to astrophysical systems, one uses, at present, two nonrelativistic, ``modified-gravity'' formulations of MOND: the aquadratic Lagrangian theory (AQUAL) \cite{bm84}, and the quasilinear MOND theory (QUMOND) \cite{milgrom10}. In addition, for the analysis of rotation curves of disc galaxies - and for their summary in what is called the mass-discrepancy-acceleration relation, or the radial-acceleration relation -- the rather generic prediction of so-called ``modified-inertia'' theories \cite{milgrom83,milgrom94,milgrom22a}, has been used, almost exclusively.
There are also various relativistic formulations of MOND, the weak-field limit of which have been used to predict gravitational lensing \cite{milgrom13,brouwer21}, and to some extent cosmology, the cosmic microwave background, and structure formation (e.g., Ref. \cite{sz21}).
For reviews of MOND see, e.g., Refs. \cite{fm12,milgrom14,milgrom20,mcgaugh20,merritt20,bz22}.
\par
However, despite the great utility and successes of these workhorses, it must be realized, as has been emphasized repeatedly, that they must only be some approximate, effective theories, with limited validity. To quote from Ref. \cite{milgrom15}: ``These theories
have much heuristic value as proofs of various concepts (e.g., that covariant MOND theories can be
written with correct gravitational lensing). But, probably, none points to the final MOND theory.
At best, they are effective theories of limited applicability. I argue that we have so far explored only
a small corner of the space of possible MOND theories.''
\par
One strong reason for this view is the appearance, in all existing theories, of an ``interpolating function'' (IF) that is introduced ``by hand'' already at the level of the action of the theory, and that artificially interpolates between a standard action at high acceleration and a deep-MOND action for low accelerations. In theories such as AQUAL and QUMOND, a single IF of a single variable is invoked. So, per force, it is this same IF that controls all phenomena -- the field of spherical systems, rotation curves of disc galaxies, the MOND external-field effect (EFE), etc.
\par
This will surely not be the case in a more fundamental MOND theory -- a FUNDAMOND.
Quantum mechanics and relativity, also make predictions that involve various IFs. But these differ from phenomenon to phenomenon, and are not introduced at the fundamental level of the theory. We expect that also in a FUNDAMOND theory there will be different IFs appearing in different contexts, and they will all be derivable from the theory and not appear as some fundamental function of the theory.
\par
In fact, even at the level of existing effective MOND theories it is not always the case that all phenomena are controlled by the same IF. For example, in the type of ``modified-inertia'' theories discussed in Refs. \cite{milgrom94,milgrom22a}, an IF appears that is specific to  circular orbits in an axisymmetric potential -- relevant for rotation curves of disc galaxies -- but that should not be relevant for other phenomena, such as for noncircular orbits, or the EFE.
\par
As we continue to search for a FUNDAMOND, we should also follow a parallel strategy in trying to construct effective theories, different from the ones we have already, even if they do not necessarily appear more fundamental.
\par
Such theories can be useful in several ways: (a) They may each point to a different avenue to the construction of a more fundamental theory. (b) Since they differ on secondary MOND predictions they may point to different observational tests that might prefer some to the others. From these we may get some idea as to which predictions of the various known MOND theories are robust, and would follow, at least approximately, in any MOND theory, and which are not. A detailed discussion of this question is to be found in Ref. \cite{milgrom14a}.
(c) Importantly, they teach us to avoid a tunnel-vision approach to MOND, and not equate MOND, at large, with some specific formulation that we happen to be using now, such as AQUAL or QUMOND. There are, in fact, relativistic formulations that give nonrelativistic MOND phenomenology, but do not lead to QUMOND or AQUAL in the nonrelativistic limit; e.g., the nonlocal theories of Refs. \cite{deffayet11,deffayet14}, the MOND adaptation of Galileon-k-mouflage theory \cite{babichev11}, and the Aether-scalar-tensor theory (AeST) \cite{sz21,verwayen23}.
\par
Such a strategy, of studying different possible extensions when looking for an underlying fundamental basis of an existing, successful theory, is common in physics, and for exactly the reasons listed above. We see this, for example, in the attempts to establish the standard model of particle physics on more fundamental grounds.
\par
In the context of MOND, one such avenue is to develop ``modified-inertia'' formulations, as is done, e.g., in Ref. \cite{milgrom22a}. These already show substantial departures from AQUAL and QUMOND in making some lower-tier predictions of MOND, as regards, e.g., the effects of the Galactic field on the
dynamics of the Solar System, or the MOND EFE.
\par
Partly with all this in mind, I describe here a class of generalizations of QUMOND that can, indeed, differ substantially from QUMOND itself on secondary predictions. They demonstrate that such diversity of predictions may occur even within the framework of modified gravity.
\par
More general two-potential theories -- generalizing AQUAL and QUMOND -- have been described in Ref. \cite{milgrom10} (Sec. 2.2). But their consequences have not been explored, as they are not so amenable to analytic investigation. In contrast, the class of theories described here are
so, which enhances their heuristic value.
\par
A class of tripotential MOND theories have been proposed and discussed in Ref. \cite{milgrom23b}.
\par
In Sec. \ref{general}, I recap QUMOND briefly, and present the generalizations in point, with some of their general properties. In Sec. \ref{examples}, I describe some examples of such generalizations, and discuss one of the examples in some more detail, discussing some of its predictions in Sec. \ref{implications}. Section \ref{discussion} is a discussion of further points.

\section{QUMOND and generalization \label{general}}
\subsection{QUMOND recap}
QUMOND is a nonrelativistic MOND formulation that replaces the Poisson equation for determining the MOND potential, $\f$, from the mass-density distribution $\r$. It invokes an auxiliary gravitational potential, $\psi$, and its Lagrangian density is
 \beq \L=-\frac{1}{\epg}\{2\gf\cdot\gps-\az\^2\Q[(\gps)\^2/\az\^2]\}  +\r(\oot\vv^2-\f).  \eeqno{i}
 The density $\r$ may be viewed as made up of the masses of the
constituents $\r(\vr,t)=\sum_i m_i\d^3[\vr-\vr_i(t)]$, treated each as a test mass in the field of the rest. $\az$ is the MOND acceleration constant; its normalization is defined such that the quintessential MOND prediction of the mass-asymptotic-rotational-speed relation (underlying the baryonic Tully-Fisher relation) is $MG\az=V^4\_\infty$.
The Lagrangian itself is then
\beq L=\int \L\drt.  \eeqno{mayate}
\par
Variation  over the particle degrees of freedom gives
 \beq \ddot\vr_i=-\gf(\vr_i),   \eeqno{ii}
 which identifies $\f$ as the (MOND) gravitational potential, as it dictates particle accelerations.
 Varying over $\f$ gives
 \beq \Delta\psi=\fpg\r, \eeqno{iia}
so, $\psi$ turns out to equal the Newtonian potential, since it satisfies the Poisson equation, sourced by $\r$ (and we supplement it by requiring $\gps\rar 0$ at infinity).
Varying over $\psi$ gives
 \beq \Delta\f=\fpg\rh\equiv\div[\n(|\gps|/\az)\gps], \eeqno{iii}
where $\n(Y)\equiv \Q'(Y^2)$.
\par
The usefulness of QUMOND is in that it prescribes the Poisson equation also for the MOND potential, only with the effective density
$\rh$ as a source.
Thus, while the theory is still nonlinear,  as a MOND theory has to be, its solution requires solving only linear differential equations, twice, with the solution of the first stage, $\psi$, feeding the source of the second stage.
Thus $\hr$ would be the deduced density if we interpret MOND accelerations in terms of Newtonian dynamics, and $\rp\equiv\hr-\r$, the so-called ``phantom mass density'' introduced in
Ref. \cite{milgrom86a}, will be interpreted as the density of ``dark matter.''
\par
The Newtonian limit of MOND requires that for $Z\rar\infty$, $Q(Z)\rar Z$, and in the opposite limit, $\n(Y)\rar Y\^{-1/2}$, or $Q(Z)\rar (4/3)Z^{3/4}$ (the normalization is set by that of $\az$).

\subsection{QUMOND generalizations}
I am here pointing out that the amenability of QUMOND to solution is retained in generalizations, whereby $Q$ is replaced by any functional of the potential $\psi$.
\par
While this does not necessarily make such theories more appealing from the fundamental point of view, it does make them quite valuable as heuristic tools.
\par
In principle, the Lagrangian density can be taken as any (possibly nonlocal) functional of $\psi$, but here I confine myself to functions of a finite number of its derivatives. We then write the Lagrangian density of this class in the form
\beq \L=\L\_G +\r({1\over 2}\vv^2-\f),~~~~~~\L\_G\equiv -\frac{1}{\epg}\{2\gf\cdot\gps-\az^2\P[\psi,\nabla\psi,\nabla^2\psi,...,\nabla^n\psi]\},   \eeqno{v}
 where $\L\_G$ is the ``free'' gravitational Lagrangian density, and $\P$, like $\Q$, is a dimensionless scalar.
\par
In such theories, $\psi$ still satisfies the Poison equation, sourced by $\r$, and is thus the Newtonian potential, and $\f$ still obeys a Poisson equation, but sourced by the effective density
\beq  \rh=\frac{\az\^2}{\epg}\nabla\cdot[\sum\_{k=0}\^n (-1)\^{k-1}\nabla^{k-1}\frac{\partial \P}{\partial \nabla^k\psi}].  \eeqno{vi}
\par
A systematic study of such extensions is far beyond the scope of this paper. So I will only discuss some examples.
But before presenting these examples, I mention several generalities concerning these theories.
\par
If $\P$ depends only on derivatives of $\psi$ of various orders, $\hr$
is a divergence field. (The $k=0$ term in Eq. (\ref{vi}) is $\propto \partial\P/\partial\psi$, and is not a divergence.)
This means that the effective mass in a volume $v$, surrounded by a surface $\S$, can be written as an integral over $\S$,
\beq    \hat M=\int\_v\hr~\drt= \frac{\az\^2}{\epg}\int\_\S~d\s\cdot \sum\_{k=1}\^n (-1)\^{k-1}\nabla^{k-1}\frac{\partial \P}{\partial \nabla^k\psi}, \eeqno{vii}
and it thus depends only on Newtonian potential and its derivatives on $\Sigma$.
\par
We see already from this expression an important departure from QUMOND: Here, even for a configuration of one-dimensional symmetry, the MOND acceleration is not a universal algebraic function of the Newtonian acceleration. This is because $\P$ itself does not depend only on the Newtonian acceleration, as is the case in QUMOND.
\par
Unlike $\gps$, higher derivatives of $\psi$ have dimensions other than acceleration. Including scalars that involve higher derivatives may permit us, indeed may compel us, to introduce dimensioned constants other than $\az$.
\par
For example, the scalar $\psi\_{,i,j}\psi\_{,i,j}$ has dimensions of (frequency)$^4$ (sum over repeated indices), $(\gps)\^2/\psi\_{,i,j}\psi\_{,i,j}$ has dimensions of (length)$^2$,
 $(\gps)^2/(\psi\_{,i,j}\psi\_{,i,j})^{1/2}$ has dimensions of a Newtonian potential (squared velocity), etc.
\par
A minimalist might consider as more appealing MOND theories that involve only $\az$ as the dimensioned constant. He can then confine himself to dimensionless scalars -- such as $\psi\_{,k}\Delta\psi\_{,k}/\psi\_{,i,j}\psi\_{,i,j}$ -- or to scalars with the dimensions of powers of acceleration, such as $\psi\_{,i}\psi\_{,i,j}\psi\_{,j}/(\psi\_{,l,m}\psi\_{,l,m})^{1/2}$, $\psi\_{,i}\psi\_{,i,j}\psi\_{,j,k}\psi\_{,k}/\psi\_{,l,m}\psi\_{,l,m}$, or
$\psi\psi\_{,i}\psi\_{,i,j}\psi\_{,j}$.
\par
But, we must allow for the possibility that other constants also appear.
For example, the nonlocal  \cite{deffayet11,deffayet14}, the Galileon-k-mouflage MOND adaptation \cite{babichev11}, and the AeST \cite{sz21,verwayen23}, relativistic formulations of MOND do contain such additional dimensioned constants.
\par
Since we view MOND, as it is now understood, as an effective, approximate theory, with $\az$ itself possibly being an effective constant that will be derived from a more fundamental theory, we must allow for the possibility that such theories might introduce additional dimensioned constants.
\par
Since we can use $\az$ (with $c$) to define constants with dimensions of length and time, we can introduce, instead of new dimensioned constant, dimensionless constants. But these have to be very different from unity, to serve our purposes (see Sec. \ref{discussion}).
\par
In theories that do introduce additional dimensioned constants, it is a convenient restriction to require that they are, nonetheless, anchored in QUMOND itself, in the sense that they satisfy the following requirement:
In regions where all the higher derivatives vanish identically, such as for a constant acceleration field\footnote{In Sec. \ref{azonly}, I discusses theories where $\az$ is retained as the only dimensioned constat. These may require a separate treatment; see there.},
\beq \P(\gps,0,...,0)=\Q[(\gps)\^2/\az\^2], ~~~~~~\frac{\partial\P}{\partial\nabla^k\psi}(\gps,0,...,0)=0,~~~ {\rm for}~~k>1. \eeqno{upsas}
\par
For the sake of completeness, I have formulated the problem for the case of any number of derivatives appearing. But we have to be wary of the fact that unlike, $\gps$, which is continuous over density discontinuity, and is discontinuous only across thin massive mass layers, higher derivatives are discontinuous across density discontinuities -- the higher the derivative, the stronger the discontinuity -- and more generally may lead to unwanted effects even where
density gradients are not infinite.
I shall thus confine myself, in all that follows, to Lagrangians depending only on the first and second derivatives, where such discontinuities, when present, are tractable, or can be made so (see Sec. \ref{caveats}).
\subsection{Forces and the center-of-mass acceleration  \label{forces}}
It was shown in Ref. \cite{milgrom14b} that for a theory of the form Eq. (\ref{v}), the force on a body that comprises the density within a volume $v$ is given by\footnote{This is obvious for a linear theory, but it holds for the present nonlinear theories as well.}
 \beq \vF_v=-\int_v\r\gf~\drt. \eeqno{tyte}
 It was also shown there that the force can be written as a surface integral over the boundary, $\S$, of $v$:
 \beq \vF_v=\int_v \div\TP~\drt=\int\_\S\TP\cdot \vec{d\s}. \eeqno{retret}
Here, $\TP$ is the stress tensor of the gravitational field, and is gotten from the gravitational Lagrangian as follows:
Using a covariantized version, $L\_G^c$, of $L\_G$, by introducing a metric, $g\_{ij}$, with all derivatives replaced by covariant derivatives, with $\drt\rar \gh\drt$, and in index contractions, $\d\_{ij}\rar g\_{ij}$, etc. Then, $\TP$ is defined such that under an infinitesimal change $g\_{ij}\rar g\_{ij} +\d g\_{ij}$ ($\d g\_{ij}$ vanishes fast enough at infinity), we have
 \beq \d L\_G^c\equiv\frac{1}{2}\int \gh \Pss^{ij}\d g\_{ij}~\drt. \eeqno{lipo}
After the covariant form of $\TP$ is gotten, we take back $g\_{ij}\rar \d\_{ij}$ to get $\TP$ itself.
\par
This was done in Ref. \cite{milgrom10} for QUMOND, and I shall not repeat it here for GQUMOND. It is, however, useful to know that the force on some mass can be written as a surface integral over any surface that encloses this mass alone.
Take any mass, $M$, in a constant Newtonian, external acceleration field, $\vg\_N\^0$. From the requirement Eq. (\ref{upsas}), all GQUMOND theories give the same constant MOND field, $\vg\^0$ (because for a constant field GQUMOND reduces to QUMOND). The mass is then subject to a MOND force that can be calculated on a large surface on which we can neglect all the higher derivatives of $\psi$ compared with the first. But, by requirement Eq. (\ref{upsas}), the combined field there is that given by QUMOND; so irrespective of the structure and internal dynamics of the mass, the force is that dictated by QUMOND, which is $M\vg\^0$.
\section{Examples\label{examples}}
I now describe some examples of GQUMOND theories.
I first discuss three (rather restricted) examples that introduce new dimensioned constants, one with screening of MOND effects in small systems, one with screening in systems with short dynamical times, and one with screening that hinge on a scalar with dimensions of gravitational potential. These are the real game changers in the present context.
I discuss only the first example in some detail; the analysis of the other two is the same {\it mutatis mutandis}.
I then discuss an example where $\az$ remains the only dimensioned constant, where the departure from QUMOND is milder.
\par
In the first three examples, $\P$ takes the specific form:
\beq   \P=f\Q[(\gps)\^2/f\az\^2], \eeqno{gatar}
where, $\Q$ has the properties it has in QUMOND. Also, $f(u)$ is a function of a single scalar variable constructed from the two first derivatives of $\psi$, with $f(u)$ monotonic, and $0< f(u)\le 1$.
\par
Since $\az$ and $f$ appear in the Lagrangian only in the product $\azs f(u)$, what this theory does, by design, is to make the effective value of $\az$ renormalized to $\az f\^{1/2}$, and it becomes dependent on properties of the system through $f(u)$.\footnote{This is reminiscent of the constants of a theory being dependent on scale, and flowing as the scale changes, in renormalization-group approaches.}
Since the normalization of $\az$ is fixed by the $MG\az=V^4\_\infty$ relation, and since we want its effective value, $\az  f\^{1/2}$, to go to this value for large $u$ -- to retain the successful MOND predictions for large $u$ -- we insist on $f$ going to 1 in this limit. If we want complete restoration of Newtonian dynamics for $u\rar 0$, whatever the acceleration is, we want $f\rar 0$ in this limit.
\par
Since $f\le 1$, by and large, such theories reduce MOND effects relative to QUMOND.
\par
Substituting this ansatz in Eq.(\ref{vi}), we get for the effective density
$$\rh=\frac{\azs}{\epg}\left[\nabla_i\left(\frac{\partial\P}{\partial\psi\_{,i}}\right)- \nabla_i\nabla_j\left(\frac{\partial\P}{\partial\psi\_{,i,j}} \right)   \right]= $$
\beq =\frac{1}{\fpg}\div(Q'\gps)+\frac{\azs}{\epg}\{\nabla_i[\frac{\partial f}{\partial\psi\_{,i}}\Q(1-\hat\Q)]-\nabla_i\nabla_j[\frac{\partial f}{\partial\psi\_{,i,j}}\Q(1-\hat\Q)
)],       \eeqno{rjosc}
where $\hat\Q$ is the logarithmic derivative of $\Q$. It approaches a value of $3/4$ for small arguments of $\Q$, and approaches 1, very fast, for argument values much larger than 1.
\par
The first term in Eq. (\ref{rjosc}) is what QUMOND gives, but with an effective value, $\az f\^{1/2}$.
\par
There are some interesting general limits:
Since $f\le 1$, in a system where $|\gps|\gg\az$, the argument in $\Q$ is also large; so $\Q(Z)=Z$ to a high accuracy, and $f$ drops out, and we get the Newtonian limit, irrespective of the value of $f$. Another limit is where $f\approx 1$, in which case we get QUMOND for all accelerations, while for $f\rar 0$ we get Newtonian dynamics for all accelerations.
\par
Note that screening does not appear because we introduce into the action or field equation, artificially, direct dependence on system attributes (such as size, mass, or dynamical time), the different screenings enter through the degree of freedom $\psi$ and its derivatives.

\subsection{Screening of MOND effects in small systems  \label{lengthscale}}
In this example of a GQUMOND theory, we make $\P$ depend, through $f$, on the scalar
 $(\gps)\^2/\psi\_{,i,j}\psi\_{,i,j}$, which  has dimensions of length squared.
So, we have to introduce a critical length and take
\beq f=f[(\gps)\^2/\ell\_0\^2\psi\_{,i,j}\psi\_{,i,j}],  \eeqno{scr1}
 having the limits
$f(u)\overset{u\rar \infty}{\longrightarrow} 1$ -- in order for the theory to approach QUMOND on large scales -- and $f(u)\overset{u\rar 0}{\longrightarrow} 0$ -- in order to screen MOND effects in small systems, and $\ell\_0$ is some length constant.
Clearly, the argument of $f$ is a measure of $(\ell/\ell\_0)\^2$, where $\ell$ is some measure of the local scale of variation of the Newtonian potential. For example, outside a spherical mass, or asymptotically far outside and bounded mass, where $\psi\approx -MG/r$, we have $u=r\^2/6\ell\_0\^2$.
\par
If we take, e.g., $f(u)=u/(1+u)$, the argument of $\Q$ becomes simply $Z=\az\^{-2}[(\gps)^2+\ell\_0\^2\psi\_{,i,j}\psi\_{,i,j}]$, and $\P$ is\footnote{Note that this expression is well defined when $\gps=0$ and $\psi\_{,i,j}=0$. See Sec. \ref{caveats}.}
\beq \P=\frac{(\gps)^2}{(\gps)^2+\ell\_0\^2\psi\_{,i,j}\psi\_{,i,j}}\Q\{[(\gps)^2+\ell\_0\^2\psi\_{,i,j}\psi\_{,i,j}]/\azs\}\equiv
(\gps)^2\SS\{[(\gps)^2+\ell\_0\^2\psi\_{,i,j}\psi\_{,i,j}]/\azs\}.  \eeqno{natar}
The limits are approached even faster, for example, with $f(u)=u\^n/(1+u\^n)$ ($n>1$), or $f(u)=1-\exp(-u\^n)$.
\par
Another mechanism of screening MOND effects at small distances was proposed in Ref. \cite{babichev11}. There, it is due to the ``Wainshtein mechanism,'' produced by an addition of a term in the relativistic MOND Lagrangian that brings in a constant $k$ with dimensions of (length)$^4$.
The screening length, $r\_V$, is then not $k\^{1/4}$ itself, but depends also on the mass of the system, and is $r\_V\sim k\^{1/4}(V\_{\infty}/c)$,
where $V\_{\infty}=(MG\az)^{1/4}$ is the asymptotic MOND circular speed for the mass in question.
\par
In the present example, the screening length is also not $\ell\_0$ itself, but is determined as follows:
Full restoration of Newtonian dynamics -- i.e., screening of all MOND effects --  requires that the argument of $Q$ in Eq. (\ref{natar}), and hence at least one of the two contributing terms, be $\gg 1$. Thus,
in an isolated system of size $r$ and characteristic, intrinsic acceleration $|\gps|=a$, some MOND effects are present if both terms are not very large; i.e., if $a\lesssim\az$ and $r\gtrsim \ell\_0(a/\az)$ [the second term is $\sim (\ell\_0 a/r\az)^2$]. If both terms are small, and the second dominates, the resulting MOND departure from Newtonian dynamics involves both $\az$ and $\ell\_0$.
\par
Note also, as already stated for the general case, that the MOND acceleration is not a universal function of the Newtonian acceleration for configuration of one-dimensional symmetry, as is the case in QUMOND.
There is also dependence on the size scale (in units of $\ell\_0$).
\par
In the limit of large argument of $\Q$, $Z\gg 1$, where $\Q(Z)\approx Z$, and $\hat\Q\approx 1$, the second and third terms in Eq. (\ref{rjosc}) become, respectively,
\beq \approx \frac{1}{\fpg}\div[(1-\hat\Q)\hat f\gps],~~~~~~~~ \approx \frac{1}{\fpg}\nabla_i\nabla_j\left[(1-\hat\Q)\hat f\frac{(\gps)^2\psi\_{,i,j}}{\psi\_{,k,l}\psi\_{,k,l}}\right],   \eeqno{batafer}
where
 \beq \hat f=\frac{\ell\_0\^2\psi\_{,i,j}\psi\_{,i,j}}{(\gps)^2+\ell\_0\^2\psi\_{,i,j}\psi\_{,i,j}}   \eeqno{gatew}
 is the logarithmic derivatives of $f$. These terms thus become negligible, with $\hat\Q$ approaching 1, and we get the Newtonian limit.
This happens when either $|\gps|\gg\az$, or when $\ell\_0(\psi\_{,i,j}\psi\_{,i,j})^{1/2}\gg\az$, even if $|\gps|\ll\az$.
\par
In the opposite limit, where $|\gps|\ll\az$, and $\ell\_0(\psi\_{,i,j}\psi\_{,i,j})^{1/2}\ll\az$, so that $Z\ll 1$, where $\Q(Z)\approx (4/3)Z^{3/4}$, we have
\beq \P\approx \frac{4\az\^{-3/2}(\gps)^2}{3[(\gps)^2+\ell\_0\^2\psi\_{,i,j}\psi\_{,i,j}]^{1/4}}.  \eeqno{mashmas}

\subsubsection{Some caveats  \label{caveats}}
Here I discuss two issues that require some clarification. One concerns what happens near density discontinuities, where $\psi\_{,i,j}$ is discontinuous. The other concerns the handling of mutual vanishing of $\gps$ and $\psi\_{,i,j}$ in relations such as Eq.(\ref{mashmas}).
\par
Across density discontinuities, $\psi\_{,i,j}$ is necessarily discontinuous, since by the Poisson equation $\Delta\psi$ is discontinuous. This can induce jumps, or ``delta-function'' contributions to the phantom density across the surface of density discontinuity. Because the fourth derivative of $\psi$ appears in the expression for the effective density, and the third derivative appears in the expression for effective column densities, even discontinuities in the derivative of the density can lead to finite jumps.
Such discontinuities are, however, harmless, since they involve finite column densities.
\par
Take, for example, a spherical mass whose density, otherwise smooth, drops from a finite value to zero at some radius $R_0$. From Eq. (\ref{vii}), the total effective mass in any concentric sphere smaller than $R_0$, or larger than $R_0$, is finite, since it is evaluated from quantities on the surface where all is smooth. Thus the difference which bounds the contribution at $R_0$, is also finite.
\par
Or, take a planar layer of smooth density $\r(z)=\r(-z)$ that drops to zero beyond some $z\_0$, with total column density $\S$. The total effective column density perpendicular to the slab, $\hat\S$, is seen, from Eq.(\ref{vi}), to depend on $\psi$ in the region outside the slab, where $|\gps|=\tpg\S$, and $\psi\_{,i,j}=0$. We then get for $\hat\S$ the QUMOND value, $\hat\S=\S\Q'[(\tpg\S/\az)^2]$ [from Eq. (\ref{upsas})]. Since the contribution to $\hat\S$ from the interior is finite, so must be that from the discontinuity.
\par
If we look down into the microscopy of the discontinuity, or in other words, if we start from a smooth transition and let it tend to a jump, we see that the effective density does diverge in a smaller and smaller region, but the leading singularity is a double layer of equal positive and negative effective densities, which does not contribute to the gravitational field outside it. In essence, if $s$ is the coordinate perpendicular to the surface of discontinuity, which is at $s=0$, the leading singularity becomes proportional to $\d'(s)$ in the limit, and $\int  \d'(s)ds=0$.
The remaining singularity is $\propto \d(s)$, leaving a finite column density.
\par
Note though that such peculiarities are not encountered on scales much larger than $\ell\_0$ -- where the effects of the second derivative can be made to disappear quickly  -- or when $|\gps|\gg\az$ -- where we are in the Newtonian limit. Such effects can appear for small systems deep in the MOND regime.
\par
I cannot think, at present of such realistic systems, but this requires further study.
\par
The other caveat concerns points, or whole regions where,
$\gps=0$ and $\psi\_{,i,j}=0$. For example, a look at Eq. (\ref{mashmas}) tells us that we need to explain what happens in such instances.
This can only occur in vacuum, since $\psi\_{,i,j}=0$ implies that $\Delta\psi=0$.
\par
Far outside a bounded mass both quantities tend to zero, but clearly, there $\psi\_{,i,j}$ vanishes faster, and the field equations tend to those of QUMOND.
\par
At a finite point, in vacuum, the Newtonian potential is analytic in the coordinates. If $\gps=0$ and $\psi\_{,i,j}=0$ at an isolated point, expanding the potential in the coordinates near this point, we see that $\gps$ vanishes with a higher power of the coordinates than $\psi\_{,i,j}$; so the limit is well defined.
\par
If both quantities vanish in a finite volume, $v$, such as in a spherical hollow in a spherical system -- or in an overlap of several such hollows -- it is still true that $\gps$ vanishes faster than $\psi\_{,i,j}$:
Beyond the boundary of $v$, we must have a finite density. Say that as a function of the distance, $\z$, from the boundary, $\r(z)\propto \z\^\a$, at some point on the boundary. Then, approaching this region from the outside, Gauss's theorem tells us that $(\gps)^2$ vanishes as $(\r\z)\^2\propto\z\^{2(\a+1)}$, while $\psi\_{,i,j}\psi\_{,i,j}$ vanishes at most like $(\Delta\psi)^2\propto\z\^{2\a}$.
\par
So we have $\P\equiv 0$ inside $v$, and the effective density vanishes identically in such a region.
\par
Seen differently, if we add a small mass somewhere, its field fully determines the effective density inside $v$; so, when we take this auxiliary extra mass to zero, the effective density in $v$ also goes to zero.

\subsection{Screening of MOND effects in systems with short dynamical times \label{timescale}}
In another example, we take the scalar argument of $f$ appearing in Eq. (\ref{gatar}) to be $\psi\_{,i,j}\psi\_{,i,j}$, which has dimensions of (frequency)$^4$; so
\beq f=f[\Om\_0\^{-4}\psi\_{,i,j}\psi\_{,i,j}],  \eeqno{scr2}
with $f(u)\overset{u\rar \infty}{\longrightarrow} 0$, and $f(u)\overset{u\rar 0}{\longrightarrow} 1$, for example,
$f(u)=(1+u\^n)^{-1}$, or $f(u)=\exp(-u\^n)$.
\par
The scalar $(\psi\_{,i,j}\psi\_{,i,j})^{1/4}$ is a measure of the local, Newtonian, dynamical frequency. For example, far outside a mass $M$ it equals $6\^{1/4}V/r$, where $V$ is the Newtonian, circular rotation speed at $r$.

\subsection{Potential-dependent screening of MOND effects \label{potescreen}}
Making the Lagrangian density depend on scalars with the dimensions of $\psi$ (or squared velocity), for example as a variable in $f(u)$, we can make the effective value of $\az$ depend on the Newtonian potential. Such a scheme was, in fact, proposed in Ref. \cite{zhao12} to remove the remaining discrepancy in galaxy clusters, even with MOND, noting that the typical gravitational potential in the cores of clusters is rather higher than in galaxies.\footnote{The suggestion of Ref. \cite{zhao12} was based on an observation by Jacob Bekenstein in a seminar he gave in 2011 at the IAP in Paris.}
Reference \cite{zhao12} started from AQUAL, and suggested to make $\az$, wherever it appears in the AQUAL Lagrangian, a function of the MOND potential.
\par
Here, I begin with QUMOND, and if we do not whish to make the Lagrangian a function of the $\psi$ itself, for reasons mentioned above, we can use other scalars with the dimensions of potential, such as $(\gps)^2/(\psi\_{,i,j}\psi\_{,i,j})^{1/2}$. This scalar has the dimensions of a gravitational potential, but it is not the local Newtonian potential. Outside a spherical mass, or asymptotically far from a bounded mass, $M$, its value is $MG/\sqrt{6}r$, but, for example, inside a spherical shell it vanishes, as explained above, and does not equal the (constant) potential there.
\par
I shall not further discuss this choice of scalars.
\subsection{Retaining $\az$ as the only dimensioned constant  \label{azonly}}
If we insist on retaining $\az$ as the only dimensioned constant, we can use in the GQUMOND Lagrangian only scalar variables that are either dimensionless, or have dimensions of acceleration to some power, which then appear as variables in $\P$ divided by the same power of $\az$.
Even though such theories depart from QUMOND less drastically, I give here an example for completeness.
\par
For the sake of concreteness, I give an example of introducing one additional variable. It demonstrates the main differences between such GQUMOND theories and QUMOND itself.
In addition to the variable $\A\equiv (\gps)^2/\azs$, which appears in QUMOND, introduce
\beq \bar A\equiv\az\^{-2}\frac{\psi\_{,i}\psi\_{,i,j}\psi\_{,j,k}\psi\_{,k}}{\psi\_{,l,m}\psi\_{,l,m}}.  \eeqno{nadta}
\par
$\P$ can then be considered a function of the two acceleration variables, or as a function of a single acceleration, e.g., $A$, and of the dimensionless ratios $\bar A/A$.
\par
The meaning of $\bar A$ is understood as follows: For a given point $\vr$, work in the principal coordinates for $\psi\_{,i,j}$, namely those for which it is diagonal at $\vr$. We then have there $\psi\_{,i,j}=diag(a\_1,a\_2,a\_3)$ ($\sum a\_i=\fpg\r$). Then,
$\bar A=\sum a\_i\^2\psi\_{,i}\^2/\sum a\_i\^2$.
This is to be compared with $(\gps)^2=\sum \psi\_{,i}\^2$. So, in $\bar A$, different components of the acceleration field $\gps$ are given different weights according to the relative eigenvalues of $\psi\_{,i,j}$.
\par
There are configurations in which the different scalars can be rather different from each other; e.g., in anisotropic ones, such as disc configuration, or low-multiplicity N-body  systems.
\par
In such theories, the single interpolating function of a single variable, as appears in QUMOND, is thus replaced with a more complicated interpolation scheme between the Newtonian and MOND regimes involving an IF of two variables.
\par
To reproduce the salient MOND predictions, this version of GQUMOND has to be space-time scale invariant in the deep-MOND limit $\az\rar\infty$ (i.e., $A,~\bar A\rar 0$). This requires the $\P(A,\bar A)$ to become homogeneous of degree 3/4. Namely, in this limit $\P$ is such that
$\P(\l A,\l\bar A)=\l\^{3/4}\P(A,\bar A)$. This, requirement is satisfied for any function of the form $\P(A,\bar A)=A^{3/4}\bar\SS(\bar A/A)$, for arbitrary $\bar\SS$.
\par
In spherical geometry where the Newtonian potential is $\psi(r)$, we have
\beq \bar A/A=\frac{(\psi'')^2}{(\psi'')^2+2(\psi'/r)^2}.   \eeqno{milop}
Outside a spherical mass, or asymptotically far from any bounded mass, we then have $\bar A/A=2/3$. So $\bar\SS$ is a constant and we get the standard asymptotic MOND field -- with an appropriate normalization of $\az$ that depends on the way $\P$ depends on $\bar A$.
\par
For example, the Newtonian potential outside a Kuzmin disc is that of a point mass (placed somewhere on the disc axis and away from the disc).
So there too QUMOND gives the correct description.
\par
But, inside a mass (even if spherical) $\bar A/A$ is not a constant and the predicted MOND field is different from that in QUMOND. And, as in GQUMOND, in general, the MOND acceleration is not a universal algebraic function of the Newtonian acceleration.
\par
In the ideal cases of exact cylindrical, or plane-parallel geometries, $\bar A/A=(\psi'')^2/[(\psi'')^2+(\psi'/r)^2]$, and $\bar A/A=1$, respectively.
\par
In the latter case, strictly speaking, $\bar A$ it is not well defined outside a finite-extent, plane-parallel, infinite slab, where $\psi\_{,i,j}=0$. However, we view the system as a limit with finite density everywhere, and with the density outside some plane vanishing, keeping the plane-parallel geometry in the limiting process, in which case $\bar A/A=1$ all along.  Since the variables in $\P$ are constant in this region according to this convention, the effective density that sources the MOND potential vanishes in such a region.
\par
Such a scale-invariant GQUMOND theory automatically satisfies the salient MOND predictions, such as asymptotic flatness of rotation curves of isolated galaxies, and the mass-asymptotic-speed relation.
\par
It was also shown in Ref. \cite{milgrom14b} that scale-invariant theories, derived from a Lagrangian density of the form given in Eq. (\ref{v}),
satisfy the same virial relation that holds in QUMOND and AQUAL: For a deep-MOND, self-gravitating, isolated system of pointlike masses, $m_p$, at positions $\vr_p$, subject to gravitational forces $\vF_p$ we have
\beq \sum_p \vr_p\cdot\vF_p=-(2/3)(G\az)^{1/2}[(\sum_p m_p)^{3/2}-\sum_p m_p^{3/2}].  \eeqno{galama}
This then leads to the same deep-MOND two-body force for arbitrary masses,
 \beq F(m_1,m_2,\ell)=\frac{2}{3}\frac{(\az G)^{1/2}}{\ell}[(m_1+m_2)^{3/2}-m_1^{3/2}-m_2^{3/2}] \eeqno{shasa}
($\ell$ is the distance between the masses),
and the same mass-velocity-dispersion relation
\beq \s^2=\frac{2}{3}(MG\az)^{1/2}[1-\sum_p (m_p/M)^{3/2}],  \eeqno{shisa}
($M=\sum_p m_p$) as in QUMOND and AQUAL.
\par
Note that as discussed in Sec. \ref{caveats}, here too the expression of $\bar A$ is well defined at isolated points where both $\gps=0$ and $\psi\_{,i,j}=0$ (where $\bar A=0$), or in extended regions where this is so, and where the resulting effective density then vanishes.
\par
However,  here, there may be an issue where $\psi\_{,i,j}=0$, but $\gps\not=0$, for example, in the ideal case where $\gps$ is exactly constant.
Since $\bar A\le A$, by definition, $\bar A$ cannot diverge in such limiting cases, but its value may not be well defined at such points.
This needs further checking. But it is clear that if such a point is isolated it might contribute a finite mass at the point [because from Eq. (\ref{vii}), the effective mass can be calculated from an integral over a surface where all quantities are well defined. This finite mass might in fact vanish. As discussed above, $\bar A$ is also not well defined for the case of an extended region where $\gps$ is exactly constant, such as outside an infinite plane-parallel slab. But such are unrealistic, and do not afford an acceptable solution even in general relativity, since it does not allow an asymptotically Minkowskian solution. And, we can self-consistently define the effective density to be zero in such regions.
\par
Note that in the present version of GQUMOND, condition (\ref{upsas}) cannot be directly applied; so the correct center-of-mass motion, which relied on this condition for the theories above (as detailed in Sec. \ref{forces}), needs to be checked separately.
At any rate, in the deep-MOND limit, the above virial relation with the two-body force, implies that the MOND force in many-pointlike-body systems, are independent of the internal structure of the bodies.
\par
Much remains to be checked and studied about such theories, to make sure that they are, in fact, self-consistent, and to deduce how their predictions depend on the choice of the interpolation scheme as defined by $\P(A,\bar A)$.
For example, one interesting version can make the dependence of $\P$ on $\bar A$ such that this dependence disappears altogether in the deep-MOND limit. In this case the theory coincides with QUMOND in the deep-MOND limit (and in the high-acceleration limit\footnote{Since, by definition, $\bar A\le A$, we have that $\bar A\gg 1\Rightarrow A\gg 1$.}), but departs from it only in the transition region, where $\bar A\sim\az$.

\section{Possible implications for astrophysical systems and phenomena \label{implications}}
For concreteness' sake I will discuss implications of the GQUMOND version that departs from QUMOND on short length scale, as defined by Eqs. (\ref{gatar})(\ref{scr1}). The implications of other versions that introduce new dimensioned constants can be treated analogously.
The versions that retain $\az$ as the only dimensioned constant require a different treatment, which I leave for the future.
\par
When dealing with an isolated system, the system length scale that enters, and that has to be compared with $\ell\_0$, is, by and large, defined by the size of the system.
When, however, we deal with a small subsystem in the field of a larger mother system, a combination of the scales of both can enter, depending on the phenomenon in question.
\par
Examples of ``subgalactic'' phenomena that may behave differently in GQUMOND and QUMOND are: the internal dynamics of small subsystems, such as binary stars and star clusters; small regions where $|\gps|\lesssim\az$ in systems whose global characteristic $|\gps|\ggg\az$, such as saddle points of star-planets, or in binary stellar systems; the small effects of a weak external acceleration field on the internal dynamics in a high-acceleration system, such as the effect of the galactic field on the dynamics of the inner Solar System; tidal effects on small systems; the dynamics perpendicular to a galactic disc; and the general MOND EFE.
I discuss some of these below.
\par
Clearly, it has to be assumed that $\ell\_0$ is small compared with the scale of galaxies, so as to retain the successes of MOND on these scales.
Below I give some important examples concerning some subgalactic systems and phenomena.
\par
A departure from QUMOND may also occur in theories that do not introduce new constants; but estimating such differences -- which are not as drastic -- is then more difficult, and depends on the exact theory.

\subsection{Isolated systems  \label{isolated}}
As explained above, in such a GQUMOND theory, MOND effects can be greatly suppressed in systems whose size is smaller than $\ell\_0$. Depending on the value of $\ell\_0$ this can suppress MOND effects to various degrees in the global internal dynamics of such systems as binary, or multiple, stellar systems, as well as open and globular star clusters, and possibly dwarf galaxies -- even without the additional quenching effect of the external galactic field.
\par
While this is obvious for the global dynamics of such systems -- as expressed, e.g., in the relation between mean velocity dispersion, size, and mass -- such departures from QUMOND can appear also in limited regions of systems that are otherwise well described by QUMOND.
\par
As an example, consider the region near the midpoint (taken as the origin) between two equal (pointlike) masses, $M$, a distance $2\ell$ apart, along the $x$ - axis.
To lowest order in the distance to the midpoint, we have $(\gps)\^2\approx 4(MG/\ell\^2)^2\ell\^{-2}(4x\^2+y\^2+z\^2)$, and $\ell\_0\^2\psi\_{,i,j}\psi\_{,i,j}\approx 24(MG/\ell\^2)^2(\ell\_0/\ell)\^2$. Suppose that system is globally in the low-acceleration regime, namely that $MG/\ell\^2\ll\az$, and that $\ell\not\ll\ell\_0$, so the higher-order term is also sub $\az$, so  Eq. (\ref{mashmas}) holds.
 At distances from the midpoint that are small relative to $\ell\_0$, the latter term dominates over the former, and we have
\beq \P\approx \frac{4\az\^{-3/2}(\gps)^2}{3(\ell\_0\^2\psi\_{,i,j}\psi\_{,i,j})^{1/4}}\approx q(\gps/\az)^2,  \eeqno{mashgis}
where
\beq  q=(2/3)^{5/4}(MG/\az\ell\^2)^{-1/2}(\ell/\ell\_0)^{1/2}.\eeqno{namana}
This is to be compared with the QUMOND expression $\P\approx (4/3)(\gps/\az)^{3/4}$.
The GQUMOND expression is Newtonian (with renormalized effective $G$) and gives rise to no phantom density in this region, whereas the QUMOND
expression gives rise to phantom matter with both positive and negative densities \cite{milgrom86a}.
\subsection{MOND effects near a sun-planet saddle point}
The possibility has been proposed and discussed (e.g., in Refs. \cite{magueijo06,bz22}) that MOND effects could be detected around points in the Solar System where the net gravitational field vanishes. There are such regions near the points where the solar field balances that of a planet (shifted somewhat by the effect of the fields of other bodies). This occurs at a distance $r\approx (m/\msun)\^{1/2}R$ from the planet (where $m$ is the mass of the planet, and $R$ its distance from the sun). In the region of interest the net $|\gps|\lesssim\az$. However, $\psi\_{,i,j}$ is strongly dominated by the contribution of the planet; so $(\ell\_0\^2\psi\_{,i,j}\psi\_{,i,j})^{1/2}\sim \ell\_0 mG/r\^3\approx (\ell\_0/r)\msun G/R^2$. Since $\msun G/R^2\gg\az$, the higher-derivative term calls the shots, and totally suppresses MOND effects, even if $r\not\ll \ell\_0$.

\subsection{Effects of the galactic field on inner Solar System dynamics}
Consider now the possible effect of the field of the galaxy in the inner Solar System, such as on planetary dynamics. It was shown \cite{milgrom09a,novak11} that in QUMOND (and in AQUAL) the external acceleration field of the Galaxy, at the solar position induces a MOND anomaly in the dynamics of the inner Solar System, which is, at present, marginally detectable \cite{hees16}.
Such a MOND effect was, in fact, claimed to explain some observed anomalies in the Solar System \cite{paucoklacka16,pauco17,migaszewski23,mathur23}
\par
The effect can be described as being due to the quadrupole field of the phantom mass that appears in the region around the point where the solar $\gps$ balances the galactic Newtonian acceleration.\footnote{In AQUAL the balance is between the MOND accelerations.} However, the higher derivatives of $\psi$ in this region are dominated by those of the field of the sun alone, since the galactic $\gps$ is nearly constant.
This means that for this phenomenon, the relevant scales appearing in the above GQUMOND examples are those characteristic of the sun at the above point of balance (a length scale of $\sim 6000$ au, and a timescale of $\sim 4.5\times 10^5 {\rm years}$). If these are smaller than $\ell\_0$ or $\Om_0\^{-1}$, respectively, then, $f(u)\ll 1$, and the MOND anomaly is greatly reduced.

\subsection{Tidal effects}
Consider a body $b$, of mass $m$, falling in the field of a mother system $B$, of mass $M$, at a distance $R$ from it (for demonstration purposes I take the two bodies to be pointlike). $b$ could be a binary star, a star cluster, or a satellite dwarf galaxy, and $B$ a mother galaxy. Or, $b$ can be a galaxy, and $B$ a cluster of galaxies. It was shown in Ref. \cite{brada00} that if we are fully in the MOND regime, then the tidal radius, $r\_t$, of $b$ due to $B$, is given essentially by the Newtonian expression.\footnote{This is because beyond the tidal radius, the external acceleration dominates over the internal; the EFE applies; the internal MOND acceleration at and beyond $r\_t$ is $\sim (MG/R^2\az)^{-1/2}(mG/r\_t\^2)$, while the increment in the MOND field of $B$ across $r\_t$ is $\sim (MG\az)^{1/2}(r\_t/R^2)$.}
Let then $\psi\_b$ be the Newtonian potential of $b$, and $\psi\_B$ that of $B$.
The tidal radius, $r\_t$, of $b$ in $B$, is then such that the differential acceleration of $B$ across an extent $r\_t$ -- which is $\sim r\_t|\nabla^2\psi\_B|\sim r\_t MG/R^3$, at the position of $b$ --  is balanced by the internal acceleration, $|\nabla\psi\_b(r\_t)|\sim mG/r\_t\^2$.
This means that around $r\_t$ we have $|\nabla^2\psi\_b|\approx|\nabla^2\psi\_B|$. Also, at $r\_t$ we have $\nabla\psi\_b\ll\nabla\psi\_B$, their ratio being of the order of $r\_t/R$. Thus, around $r\_t$, both $(\gps)\^2$ and $\psi\_{,i,j}\psi\_{,i,j}$ are determined by $B$, and the scales that enter are characteristic of $B$ alone, for which we assume that $f(u)\approx 1$, and so GQUMOND reduces to QUMOND. So, tidal effects in such versions of GQUMOND are expected to be the same as those in QUMOND.
\section{Discussion  \label{discussion}}
I have pointed out a class of nonrelativistic generalizations of the QUMOND formulation of MOND -- dubbed here ``GQUMOND'' -- whose Lagrangian density depends also on higher space derivatives of the (Newtonian) gravitational potential. Even if we confine ourselves only to first and second derivatives of the potential, this allows us to define various scalars that can appear in the Lagrangian. Many versions of GQUMOND can thus be constructed involving the various scalars in different combinations. Only a small part of the possibilities, with their possible consequences, have been considered, but the gist of the idea should be clear.
\par
Similar generalizations may be considered for AQUAL, as well. However, their mathematical structure, with AQUAL being fully nonlinear, would be rather more complex, and the theory would be rather more difficult to analyze. With GQUMOND we are on a rather safe ground; e.g., as regards the assurance of existence and uniqueness of solutions, since we are only dealing with the Poisson equation.
\par
In one GQUMOND subclass, which I bring here only for completeness, $\az$ remains the only dimensioned constant in the theory. Theories in this class can differ from QUMOND in their predictions, but the departure is not expected to be so drastic.
\par
In contradistinction, the introduction of higher derivatives can force the introduction of additional dimensioned constants, and this allows a much larger variety, and may entail substantial departures from the predictions of QUMOND, as I demonstrated  with several instances.
\par
Introducing new dimensioned constants is, however, a price to pay in terms of simplicity and economy. We can use $\az$ with $c$ to define other dimensioned constants, such as the MOND length $\ell\_M\equiv c\^2/\az$, or the MOND time $t\_M\equiv c/\az$, but these are characteristic of the Universe at large, given the well-known coincidence,
\beq \baz \equiv 2\pi \az\approx a\_H(0)\equiv cH_0\approx a\_\Lambda\equiv c^2/\ell\_{\Lambda}, \eeqno{coinc}
where $H_0$ is the value of the Hubble-Lemaitre constant, and  $\ell\_{\Lambda}=(\Lambda/3)^{-1/2}$ is the radius associated with $\Lambda$ -- the observed equivalent of a cosmological constant. Such constants are not what we need here to make a difference. So, we can replace the dimensioned constants we introduced above with some dimensionless multiples of these MOND units -- for example, take $\ell\_0=k\ell\_M$, with $k\ll 1$.
\par
However, perhaps this is what nature dictates, since, in any event, MOND as we now know it can only be an effective theory, approximating some theory, a FUNDAMOND, at a deeper level. And perhaps some version of GQUMOND is a better clue for where to look for such a FUNDAMOND.
\par
But, in the least, such GQUMOND theories are important heuristic tools in demonstrating that when it comes to secondary predictions of MOND, different formulations can make rather different predictions, and that we should not view such second-tier predictions of past MOND formulations, such as QUMOND, as absolute predictions of the MOND paradigm itself.

\end{document}